\pdfoutput=1
\documentclass[twocolumn,superscriptaddress,aps,preprintnumbers,amsmath,amssymb,prd,nofootinbib,floatfix]{revtex4-2}

\usepackage{amsmath}
\usepackage{amssymb}
\usepackage{amsfonts}
\usepackage{mathrsfs}
\usepackage{graphicx}
\usepackage{xcolor}
\usepackage{physics}
\usepackage{fourier}
\usepackage{hyperref}
\usepackage{bm}
\usepackage{booktabs}
\usepackage{placeins}

\definecolor{rossoferrari}{HTML}{D9073D}
\definecolor{lightblue}{rgb}{0.284602,0.317763,0.963947}
\hypersetup{
  setpagesize=false,
  bookmarksnumbered=true,
  bookmarksopen=true,
  colorlinks=true,
  linkcolor=lightblue,
  urlcolor=rossoferrari,
  citecolor=rossoferrari,
  linktocpage=false,
  pdftitle={Formation and Charge Relaxation of Julia-Zee Dyons in Classical Lattice Simulations},
  pdfauthor={Masaki Yamada}
}

\newcommand{\ii}{\mathrm{i}}

\begin{document}

\title{Formation and Charge Relaxation of Julia-Zee Dyons in Classical Lattice Simulations}

\author{Masaki Yamada}
\email{m.yamada@tohoku.ac.jp}
\affiliation{Department of Physics, Tohoku University, Sendai, Miyagi 980-8578, Japan}

\preprint{TU-1326}

\date{\today}


\begin{abstract}
We demonstrate the formation and subsequent evolution of Julia--Zee dyons in $(3+1)$-dimensional classical lattice simulations of an $\mathrm{SU}(2)\to\mathrm{U}(1)$ symmetry-breaking model in an expanding background.
The gauge and Higgs fields are prepared thermally, subjected to an ultraviolet filter, and evolved through a mass quench and subsequent expansion.
Classical dyons can carry continuous electric charge, and we find charge magnitudes exceeding four times the electric charge unit $g$.
Their measured electric-charge magnitudes tend to decrease during the subsequent evolution, consistent with charge exchange with the surrounding fields.
At late times, the magnetic monopoles persist while the number of dyons and their observed charge range decrease, with only a few identified dyons remaining at the end of the simulations.
\end{abstract}

\maketitle

\section{Introduction}
\label{sec:introduction}

Magnetic monopoles play important roles in both phenomenology and the formal study of gauge theories~\cite{Harvey:1996ur,Mitsou:2026zcf}.
Dirac showed that the existence of a magnetic monopole requires electric charge to be quantized~\cite{Dirac:1931kp} (see also Ref.~\cite{Hull:2024uwz} for a formulation using generalized symmetries).
The 't Hooft--Polyakov solutions subsequently established that smooth, finite-energy monopoles arise in a renormalizable $\mathrm{SU}(2)$ gauge theory broken to $\mathrm{U}(1)$ by an adjoint Higgs field~\cite{tHooft:1974kcl,Polyakov:1974ek}.
Electric-magnetic duality further relates electrically charged particles and magnetic monopoles, with monopoles appearing as elementary fields in a weakly coupled dual description in supersymmetric examples~\cite{Montonen:1977sn,Seiberg:1994rs}.
In cosmology, monopoles produced in a hidden gauge sector have been proposed as dark matter~\cite{Murayama:2009nj,Khoze:2014woa,Sato:2018nqy,Yang:2022quy,Zhou:2025zzz,Brummer:2025inh,Brummer:2026wma}.

Monopoles can form through the Kibble mechanism~\cite{Kibble:1976sj,Zurek:1985qw,Zurek:1994xs} or thermal gauge-field fluctuations~\cite{Hindmarsh:2000kd,Rajantie:2001ps,Rajantie:2002dw}.
Classical simulations have explored their production, scattering, annihilation, and evolution in expanding backgrounds~\cite{Vachaspati:2016abz,Vachaspati:2015ahr,Bachmaier:2025jaz,Hindmarsh:2025vxh,Di:2026bnb,Hemming:2026osi,Hemming:2026ubo}.
Related studies address their interactions with strings and domain walls in theories with additional symmetry breaking~\cite{Langacker:1980kd,Ng:2008mp,Dvali:1997sa,Brush:2015vda,Bachmaier:2023zmq,Bachmaier:2026bgw}.\footnote{
For related studies of global monopole networks, see Refs.~\cite{Yamaguchi:2001xn,Lopez-Eiguren:2016jsy,Nakano:2026zme}, and for radiation and cosmological signatures, see Refs.~\cite{Aburatani:2026rct,Kanda:2026oon}.
Other lattice studies address the equilibrium monopole gas~\cite{Niemi:2022bjg} and instantons for monopole pair production in strong magnetic fields~\cite{Ho:2021uem}.}

A monopole carrying nonzero electric charge is called a dyon.
In the quantum theory, the electric and magnetic charges of two dyons obey the Dirac--Schwinger--Zwanziger condition~\cite{Schwinger:1968rq,Zwanziger:1968ams,Zwanziger:1968rs}. 
For the bosonic adjoint theory considered here, quantization of the monopole's electric collective coordinate gives $Q_e=g n_e$ at vanishing vacuum angle, where $g$ is the gauge coupling and $n_e\in\mathbb Z$~\cite{Tomboulis:1975qt,Witten:1979ey}.
In the presence of magnetic charge, a vacuum angle $\theta$ shifts the electric charge through the Witten effect, even in the low-energy $\mathrm{U}(1)$ description~\cite{Witten:1979ey}.
Promoting $\theta$ to an axion-dependent angle makes the dyon energy depend on the axion field~\cite{Fischler:1983sc}.
The resulting monopole-induced axion potential has been used to suppress the axion abundance~\cite{Kawasaki:2015lpf,Nakagawa:2020zjr,Banerjee:2024ykz} and isocurvature fluctuations~\cite{Nomura:2015xil,Kawasaki:2017xwt}.
An axion potential can also arise from virtual monopole loops~\cite{Fan:2021ntg}. 
Axion dissipation through dyon discharge has also been proposed, with mechanisms involving charged vector bosons~\cite{Jeong:2023gjc} or transitions between quantized dyon levels through emission of light charged fermions~\cite{Co:2025lcs}.
These effects motivate understanding how a monopole's electric charge responds to its environment.

Classical field equations already admit dyons without a vacuum-angle term.
Julia and Zee constructed static, spherically symmetric solutions in the $\mathrm{SU}(2)$ adjoint Higgs theory with both magnetic and electric charge~\cite{Julia:1975ff}.
Magnetic winding is topological, but the electric charge of an individual classical monopole is not fixed by that winding.
Its electric charge is a continuous parameter at the classical level, in contrast to the discrete spectrum obtained after quantization.
An isolated dyon does not lower its energy by emitting free charged vector bosons~\cite{Coleman:1976uk,Loginov:2021gko}.
However, this energetic stability does not prevent charge exchange with charged fields already present in its surroundings.
Such exchange occurs in low-energy charged-gauge-boson scattering from a monopole~\cite{Cvetic:1987th}.
The distribution of monopole electric charges therefore depends on the initial field fluctuations and their subsequent dynamics.

In this work, we study dyon formation and electric-charge evolution using classical lattice simulations in three spatial dimensions, with $\mathrm{SU}(2)\to\mathrm{U}(1)$ symmetry breaking in an expanding background. 
The $\mathrm{SU}(2)$ gauge fields and the adjoint Higgs field are prepared thermally, subjected to an ultraviolet filter, and evolved through a mass quench and subsequent cosmic expansion.
The simulations produce dyons with continuous electric charges and show a decrease in the observed charge range and the number of identified dyons at late times.
Tracked objects exhibit charge reduction and possible pair annihilation. 

In Sec.~\ref{sec:model}, we specify the theory and construct static Julia--Zee solutions for comparison with the lattice configurations.
Sec.~\ref{sec:numerics} describes the simulations and follows the evidence from localized charged configurations to individual charge histories and population statistics.
We discuss the implications and limitations in Sec.~\ref{sec:conclusions}.

\section{Model and Julia--Zee dyons}
\label{sec:model}

In this section, we specify the gauge theory and construct isolated Julia--Zee solutions assuming spherical symmetry.

\subsection{Model}

We consider a classical $\mathrm{SU}(2)$ gauge theory with a real adjoint scalar $\phi^a$ ($a=1,2,3$).
In units $\hbar=c=1$, its action is
\begin{align}
 S=\int d^4x\sqrt{-\det g_{\mu\nu}}\,
 \bigg[&-\frac14 F^a_{\mu\nu}F^{a\mu\nu}
 +\frac12(D_\mu\phi)^a(D^\mu\phi)^a \nonumber\\
 &-V(\phi,\eta)\bigg],
 \label{eq:model-action}\\
 V(\phi,\eta)={}&\frac12m^2(\eta)\phi^a\phi^a
                  +\frac{\lambda}{4}(\phi^a\phi^a)^2 .
 \label{eq:model-potential}
\end{align}
We use the fixed background metric $ds^2=R^2(\eta)(d\eta^2-d\bm x^2)$, where $\eta$ is conformal time and $R(\eta)$ is the scale factor.
With generators $T^a=\sigma^a/2$, we define the covariant derivative and field strength by
\begin{align}
 (D_\mu\phi)^a&=\partial_\mu\phi^a
                    -g\epsilon^{abc}A_\mu^b\phi^c,\\
 F^a_{\mu\nu}&=\partial_\mu A_\nu^a-\partial_\nu A_\mu^a
                    -g\epsilon^{abc}A_\mu^b A_\nu^c.
 \label{eq:continuum-conventions}
\end{align}
These signs agree with the link convention $U_{x,i}=\exp[\ii g\Delta x A_i^a T^a+O((\Delta x)^2)]$ and adjoint transport used below.
During the quench in the lattice simulation, the mass parameter $m^2(\eta)$ changes from $m_i^2>0$ to $m_f^2<0$ and remains fixed thereafter.
The final potential, $V(\phi) \equiv V(\phi, \eta)$ at sufficiently late times, gives
\begin{align}
 v^2&=-\frac{m_f^2}{\lambda},&
 V(\phi)-V(v)&=\frac{\lambda}{4}(\phi^a\phi^a-v^2)^2,
 \label{eq:broken-potential}\\
 m_W&=gv,&m_H&=\sqrt{2\lambda}\,v .
 \label{eq:broken-scales}
\end{align}
Here $v$ is the Higgs vacuum expectation value, and $m_W$ and $m_H$ are the charged vector and Higgs masses, respectively.
The vacuum leaves an unbroken $\mathrm{U}(1)$ gauge field.
The vacuum manifold $\mathrm{SU}(2)/\mathrm{U}(1)\simeq S^2$ supports integer magnetic winding.
The theory also admits smooth, finite-energy monopoles carrying electric charge, known as Julia--Zee dyons~\cite{Julia:1975ff}.
Their electric charge is continuous in the classical theory.

Throughout this paper, we express energies and physical lengths in units of $\sqrt{-m_f^2}$ and its inverse.
In these units, we set $m_f^2 = -1$. 
We use $g=1$ and $\lambda=0.25$, as in the simulations, so that $v=2$, $m_W=2$, and $m_H=\sqrt{2}$.

\subsection{Spherical ansatz, electric charge, and energy}

We construct static Julia--Zee solutions at $m^2=m_f^2$ in flat spacetime, using the physical radial coordinate $r$.
These isolated static solutions are used for comparison with the lattice configurations, which evolve through the quench and subsequent expansion.
For further studies of monopole and dyon profiles and internal structure, see Refs.~\cite{Bais:1975bq,Forgacs:2005vx,Panteleeva:2023aiz}.

For one unit of positive magnetic winding, we use the static, spherically symmetric ansatz
\begin{align}
 \phi^a&=v h(x)\hat r^a,&
 A_0^a&=-v j(x)\hat r^a,\nonumber\\
 A_i^a&=\epsilon_{iab}\hat r^b\frac{1-w(x)}{gr},&
 x&\equiv gv r,
 \label{eq:jz-ansatz}
\end{align}
where $\hat{\bm r}=\bm r/r$ is the radial unit vector, whose components are identified with internal directions in the spherically symmetric ansatz. 
This ansatz retains $A_0$ to describe the static electric field, whereas the lattice evolution is performed in temporal gauge.
Defining $\kappa \equiv \lambda/g^2$ and denoting $d/dx$ by a prime, the field equations reduce to~\cite{Julia:1975ff,Nishino:2019lqm}
\begin{align}
 w''&=w\left[\frac{w^2-1}{x^2}+h^2-j^2\right],\nonumber\\
 h''+\frac{2}{x}h'&=\frac{2w^2h}{x^2}+\kappa h(h^2-1),\nonumber\\
 j''+\frac{2}{x}j'&=\frac{2w^2j}{x^2}.
 \label{eq:jz-odes}
\end{align}
Regularity and finite energy conditions require
\begin{align}
 w(0)&=1,&h(0)&=0,&j(0)&=0,\nonumber\\
 w(\infty)&=0,&h(\infty)&=1,&j(\infty)&=\nu .
 \label{eq:jz-bcs}
\end{align}
For $\nu=0$, the electric profile vanishes, $j=0$.
Nonzero $\nu$ generates a regular electric field while retaining the magnetic winding.
We follow the branch connected to the uncharged monopole and restrict to $|\nu|<1$.

Away from zeros of $\phi$, the projected $\mathrm{U}(1)$ field strength can be defined by the gauge-invariant tensor introduced by 't Hooft~\cite{tHooft:1974kcl},
\begin{align}
 \mathcal F_{\mu\nu}&=n^aF^a_{\mu\nu}
 +\frac1g\epsilon^{abc}n^a(D_\mu n)^b(D_\nu n)^c,\nonumber\\
 n^a&\equiv \frac{\phi^a}{|\bm\phi|}.
 \label{eq:continuum-projected-field}
\end{align}
In flat spacetime, $E_i=\mathcal F_{0i}$ and $B_i=\tfrac12\epsilon_{ijk}\mathcal F_{jk}$.
With the sign of $A_0$ in Eq.~(\ref{eq:jz-ansatz}), positive $\nu$ corresponds to positive electric charge.
For the spherical solution,
\begin{align}
 E_r&=gv^2j'(x),& B_r&=\frac{1}{gr^2},\nonumber\\
 j(x)&=\nu-\frac{q}{x}+ \mathcal{O}(x^{-2}),&
 q&=\lim_{x\to\infty}x^2j'(x),\\
 Q_e&=\int_{S^2_\infty}\bm E\cdot d\bm S
       =\frac{4\pi}{g}q,& Q_m&=\frac{4\pi}{g}.
 \label{eq:jz-charges}
\end{align}
We use $Q_e$ for the electric flux, with the same normalization as the lattice charge measurements.
The parameter $q=Q_e/Q_m$ denotes the ratio of electric to magnetic charge.
The electric charge unit in this bosonic adjoint theory is the charged vector-boson charge magnitude $g$, so the charge in these units is $Q_e/g$.
Since $g=1$ throughout our numerical results, the plotted values of $Q_e$ coincide numerically with $Q_e/g$.

\begin{figure}[tb]
 \centering
 \includegraphics[width=\columnwidth]{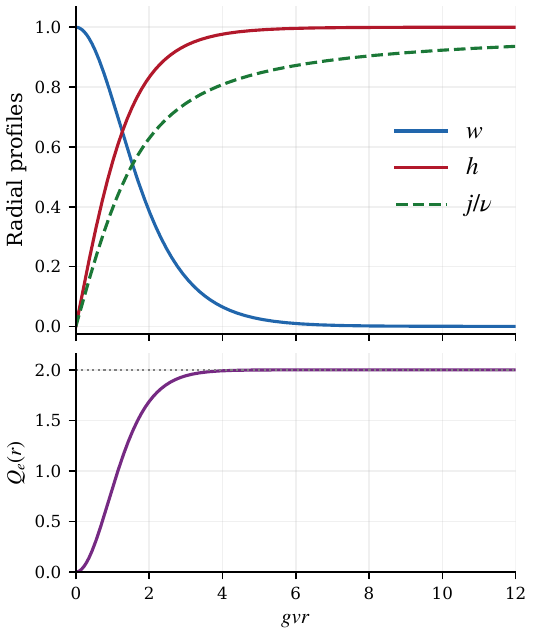}
\caption{Example of a static Julia--Zee solution with $g=1$, $\lambda=0.25$, $v=2$, and $Q_e=2$.
The upper panel shows the profiles $w$, $h$, and $j/\nu$ defined in Eq.~(\ref{eq:jz-ansatz}).
The lower panel shows the electric flux $Q_e(r)=(4\pi/g)x^2j'(x)$ through a sphere of radius $r$, where $x=gvr$.
The dotted line is its asymptotic value.
}
 \label{fig:jz-profile}
\end{figure}

The total static energy is
\begin{align}
 M&=\frac{4\pi v}{g}\,\mathcal M,\qquad
 \mathcal M=\int_0^\infty dx\,\mathcal E(x),\nonumber\\
 \mathcal E(x)&=w'^2+\frac{(w^2-1)^2}{2x^2}
       +\frac{x^2h'^2}{2}+w^2h^2\nonumber\\
   &\quad+\frac{\kappa x^2}{4}(h^2-1)^2
       +\frac{x^2j'^2}{2}+w^2j^2 .
 \label{eq:jz-energy}
\end{align}
For the chosen parameters, the vacuum energy density is $V(v)=-\lambda v^4/4=-1$.
We subtract this constant when calculating $M$, whereas the lattice-simulation plots retain it in the signed energy density.

We solve Eq.~(\ref{eq:jz-odes}) as a boundary-value problem using the adaptive collocation solver in SciPy~\cite{Virtanen:2019joe} and continuation in $\nu$.%
\footnote{
Complementary analytic constructions of monopole and dyon profiles at nonzero Higgs self-coupling using systematic asymptotic expansions have recently been developed~\cite{Malinsky:2026eux,Malinsky:2026uum,Dunne:2026hfx}.
}
The boundary conditions retain the Coulomb tail, and we add its energy outside the numerical domain analytically.
Figure~\ref{fig:jz-profile} gives an example with $Q_e=2$, $\nu\simeq0.207$, and $M\simeq30.3$.
The Higgs magnitude vanishes at the origin and approaches its vacuum value at large radii, while the gauge profile $w$ tends to zero.
The reference core length is
\begin{align}
 r_{\mathrm{core}}=\max\!\left[
 m_W^{-1},m_H^{-1}\right].
\end{align}
Outside the core, the enclosed electric flux approaches a nonzero constant.

\begin{figure}
 \centering
 \includegraphics[width=\columnwidth]{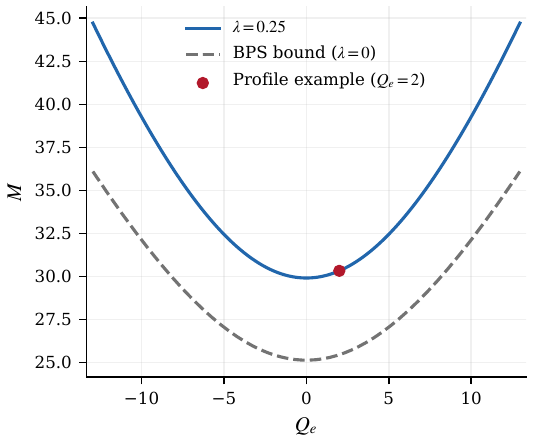}
\caption{Vacuum-subtracted static dyon energy versus electric flux at fixed $Q_m=4\pi/g$, $g=1$, and $v=2$.
The solid curve is the numerical $\lambda=0.25$ branch, with negative charges obtained by charge reversal.
The dashed curve is the BPS bound, corresponding to $\lambda=0$ at the same $v$.
The marked point is the example in Fig.~\ref{fig:jz-profile}.}
 \label{fig:jz-charge-energy}
\end{figure}

Figure~\ref{fig:jz-charge-energy} shows the energy as a function of electric charge for the same couplings with $0\leq\nu\leq0.99$.
The neutral monopole has $M\simeq29.9$, and $M$ increases with $|Q_e|$ on the sampled branch.

An analytic check is provided by the Prasad--Sommerfield limit $\lambda\to0$ at fixed $v$~\cite{Prasad:1975kr}.
With $s=\sqrt{1-\nu^2}$, its profiles are
\begin{align}
 w(x)&=\frac{sx}{\sinh(sx)},&
 h(x)&=\coth(sx)-\frac{1}{sx},\nonumber\\
 j(x)&=\nu h(x),&q&=\frac{\nu}{s},
 \label{eq:jz-bps-profiles}\\
 M_{\mathrm{BPS}}&=v\sqrt{Q_m^2+Q_e^2}
                  =\frac{4\pi v}{g s}.
 \label{eq:jz-bps}
\end{align}
The mass $M_{\mathrm{BPS}}$ gives the Bogomolny--Prasad--Sommerfield (BPS) lower bound~\cite{Bogomolny:1975de}, shown by the dashed curve in Fig.~\ref{fig:jz-charge-energy}.

\section{Lattice simulations of dyon formation}
\label{sec:numerics}
\label{sec:parameters}
\label{sec:results}

\begin{figure*}[tb]
 \centering
 \includegraphics[width=0.98\textwidth]{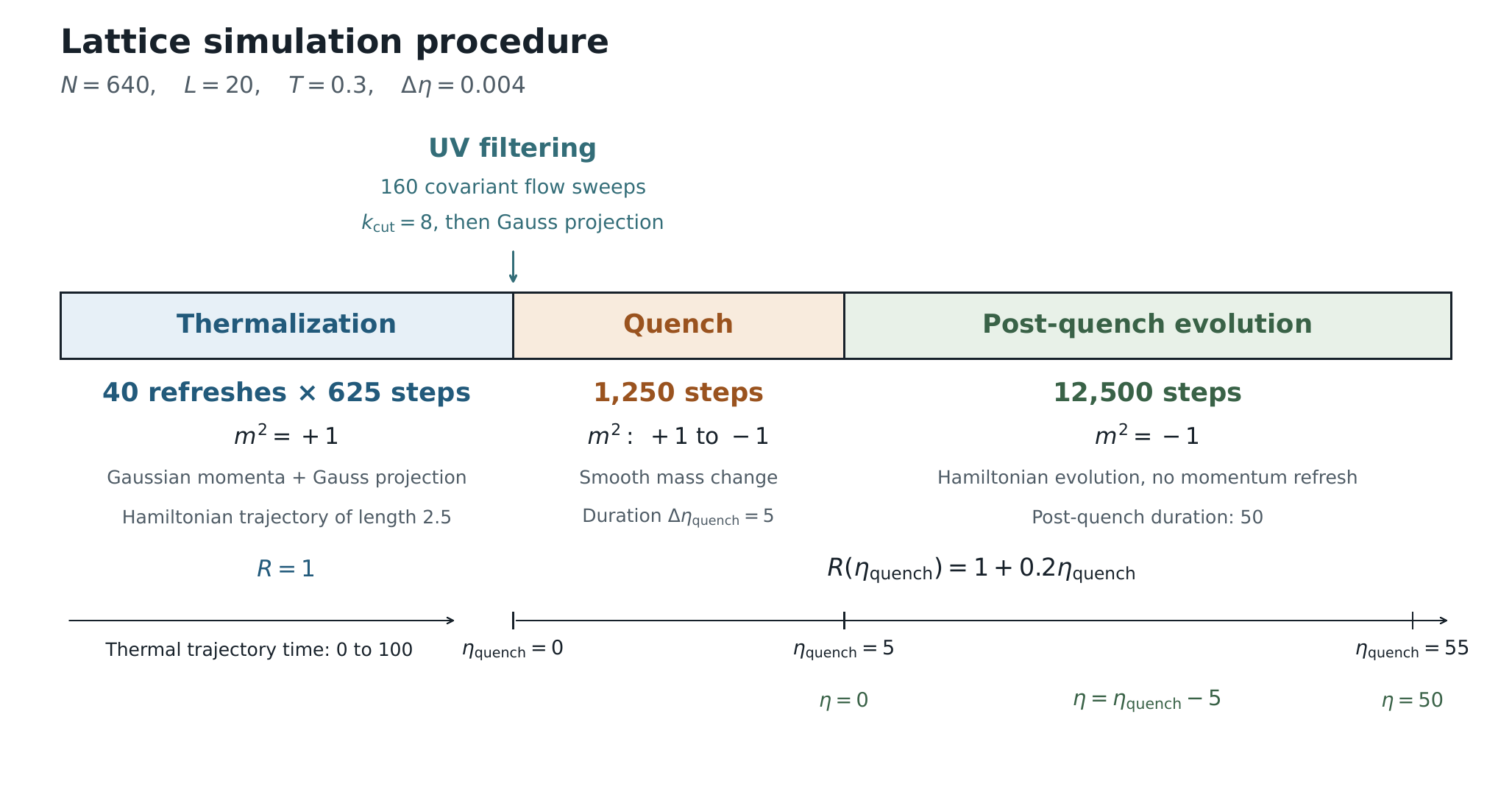}
\caption{Simulation procedure for $N=640$, $L=20$, $T=0.3$, and $\Delta\eta=0.004$.
The blue stage contains $40$ momentum refreshes, each followed by Gauss projection and a $625$-step Hamiltonian trajectory at $m^2=+1$ and $R=1$.
The UV filter applies $160$ covariant flow sweeps and a final Gauss projection before the quench.
The orange stage changes the mass parameter $m^2$ from $+1$ to $-1$ over $1250$ steps, and the green stage evolves at $m^2=-1$ for $12{,}500$ steps.
The thermal trajectory clock is separate from the conformal clock, which starts at the onset of the quench.
The quench lasts $\Delta\eta_{\mathrm{quench}}=5$, and the post-quench clock is $\eta=\eta_{\mathrm{quench}}-5$. 
The lower arrows indicate the clocks, and $R(\eta)=2+0.2\eta$ applies during the post-quench evolution.
}
 \label{fig:simulation-protocol}
\end{figure*}

Using three-dimensional lattice simulations, we follow the formation of localized magnetic defects, examine their electric fields and charge profiles, and test whether their electric charge persists.
We first specify the initial state, evolution, and charge measurements.
The results proceed from spatial evidence for dyons to individual charge histories, the charge distribution, and finally the magnetic and dyon number densities.

\subsection{Numerical methods}
\label{subsec:simulation-setup}

We evolve the classical gauge and Higgs fields on a periodic cubic lattice with $N^3$ sites, using the Hamiltonian formulation of lattice gauge theory~\cite{Kogut:1974ag}.%
\footnote{
Classical real-time lattice simulations of non-Abelian gauge and Higgs fields were used in early studies of thermal sphaleron transitions~\cite{Ambjorn:1990wn,Ambjorn:1990pu}.
}
The dynamical variables are spatial $\mathrm{SU}(2)$ links $U$ in temporal gauge, their conjugate electric momenta $P$, and the adjoint scalar $\phi$ with conjugate momentum $\Pi$.
The comoving lattice spacing is denoted by $\Delta x$, and the scale factor is $R(\eta)$.
Our main simulation uses an $N^3 = 640^3$ lattice and provides the spatial plots and individual charge histories.
The comoving side length of the simulation box is $L=20$, giving $\Delta x=0.03125$ and $\Delta\eta/\Delta x=0.128$ for $\Delta\eta=0.004$.
For the population statistics, we also use three $N=512$, $L=20$, $\Delta\eta=0.005$ runs with different seeds.
A fifth run has $N=384$, $L=10$, $\Delta\eta=0.005$.
Table~\ref{tab:parameters} gives the detailed simulation parameters used in the main calculation. 
The detailed definitions and algorithms are given in Apps.~\ref{app:preparation} and \ref{app:charge}.

The calculation proceeds through the preparation, ultraviolet (UV) filtering, mass quench, and subsequent evolution, as summarized in Fig.~\ref{fig:simulation-protocol}. 
Here we briefly describe each stage of the simulation.

To prepare the initial thermal fluctuations of the gauge and Higgs fields, we alternate Gaussian momentum refreshes with Hamiltonian trajectories while enforcing the lattice Gauss constraint.
Each refresh retains $U$ and $\phi$ and draws new $P$ and $\Pi$ with variances set by $T=0.3$ through Eq.~(\ref{eq:refresh-variances}).
We project these momenta onto the lattice Gauss constraint using a preconditioned conjugate-gradient solver before evolving the next trajectory.
For the $N=640$ run, we perform $40$ trajectories each consisting of $625$ steps over a duration of $2.5$.
No Metropolis accept/reject step is applied.

An unregulated classical thermal ensemble has a Rayleigh--Jeans ultraviolet divergence~\cite{Aarts:1999wj}.
We therefore apply a gauge-covariant ultraviolet filter before the transition, with a fixed physical smoothing scale that suppresses short-wavelength fluctuations as the lattice is refined.
Each sweep updates all links with the Wilson gauge force and then smooths $\phi$, $\Pi$, and each adjoint momentum channel $P_i$ using gauge-transported nearest-neighbor differences.
For the $N=640$ run, we apply $160$ sweeps and restore the Gauss constraint after the final sweep.
The filter's relation to the physical cutoff $k_{\mathrm{cut}}=8$ is given in App.~\ref{app:evolution}.

The subsequent Hamiltonian evolution uses a second-order velocity-Verlet integrator with a kick--drift--kick update.
Link updates use $\mathrm{SU}(2)$ exponentials, while the scalar update includes the integral of $R^{-2}$ over the time step.
The mass parameter and scale factor are evaluated at the corresponding step endpoints.
During the quench and subsequent evolution, we monitor the Gauss residual and link unitarity.

During the quench, we measure conformal time from its onset and denote it by $\eta_{\mathrm{quench}}$.
We change the mass parameter from $m_i^2=+1$ to $m_f^2=-1$ using the smooth interpolation in Eq.~(\ref{eq:mass-quench}).
The quench lasts $\Delta\eta_{\mathrm{quench}}=5$, corresponding to $1250$ steps, while $R=1+0.2\eta_{\mathrm{quench}}$ increases from $1$ to $2$.
For the subsequent evolution, we reset the time origin and use $\eta=\eta_{\mathrm{quench}}-5$.
We then evolve for $12{,}500$ steps, reaching $\eta=50$, with $m^2=m_f^2$ and $R=2+0.2\eta$.

\begin{table}[t]
\caption{Preparation and evolution parameters of the $N=640$ realization.
The thermal clock is separate from the quench and post-quench clock.
Numerical tolerances and the measurement definitions are given in Apps.~\ref{app:preparation} and \ref{app:charge}.}
 \label{tab:parameters}
 \centering
 \begin{tabular}{@{}ll@{}}
  \toprule
  Quantity & Value \\
  \midrule
  $N$, $L$, boundary conditions & $640$, $20$, periodic \\
  $\Delta x$ & $0.03125$ \\
  $\Delta\eta$, $\Delta\eta_{\mathrm{th}}$ & $0.004$, $0.004$ \\
  $g$, $\lambda$, $T$ & $1$, $0.25$, $0.3$ \\
  $m_i^2$, $m_f^2$ & $+1$, $-1$ \\
  Initial scalar Gaussian width & $0.08$ \\
  Thermal trajectory length & $2.5$ ($625$ steps) \\
  Momentum refreshes & $40$ \\
  Physical UV cutoff $k_{\mathrm{cut}}$ & $8$ \\
  Quench duration & $5$ ($1250$ steps) \\
  Quench center $\eta_c$, width $w$ & $2.25$, $0.6$ \\
  $R_0$, $h_0$ in $R=R_0(1+h_0\eta_{\mathrm{quench}})$ & $1$, $0.2$ \\
  Post-quench duration & $50$ ($12{,}500$ steps) \\
  \bottomrule
 \end{tabular}
\end{table}

\subsection{Results}
\label{subsec:spatial-formation}

To measure magnetic and electric charges in the Higgs phase, we extract the residual $\mathrm{U}(1)$ field by projecting onto the local Higgs direction, following the lattice construction of Refs.~\cite{Davis:2000kv,Edwards:2009bw}.
Details of the procedure are given in App.~\ref{app:charge}.

\begin{figure*}[!tp]
 \centering
 \includegraphics[width=\textwidth]{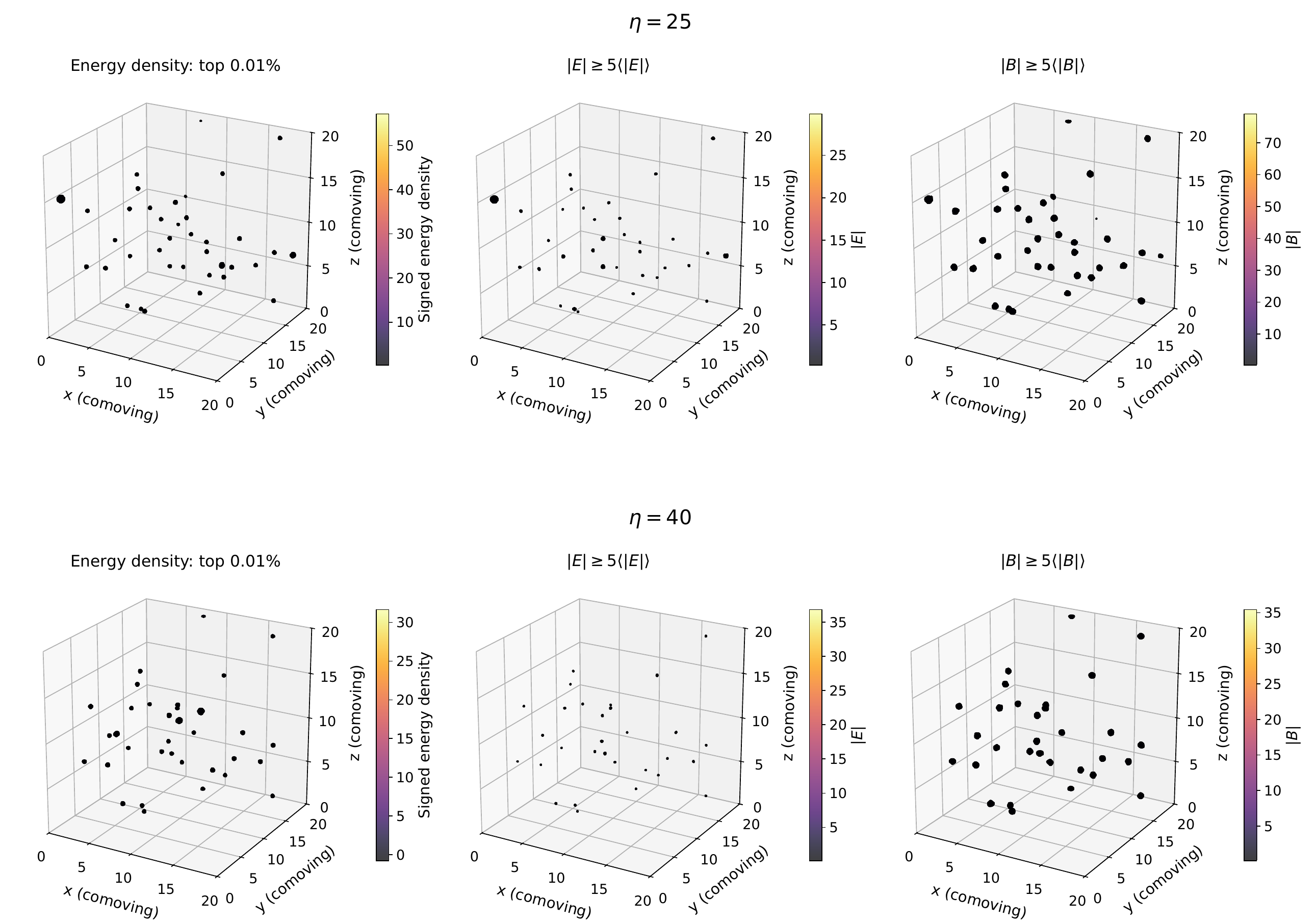}
\caption{Grid points of large field values in the $N=640$, $L=20$ run, plotted in comoving coordinates.
The upper and lower rows correspond to $\eta=25$ and $40$.
From left to right, the columns show energy density (largest $0.01\%$), $|E|\geq5\langle|E|\rangle$, and $|B|\geq5\langle|B|\rangle$.
The color bars show physical energy density and field magnitudes, and the averages are over the full box at each time.
Overlapping points can obscure the color of an interior maximum.}
 \label{fig:volume640}
\end{figure*}

We first examine the localized structures produced by symmetry breaking.
Figure~\ref{fig:volume640} shows compact concentrations of energy density $\rho$, electric-field magnitude $|E|$, and magnetic-field magnitude $|B|$ at $\eta=25$ and $40$.
The energy panels show the $0.01\%$ of grid points with the largest energy density.
The electric and magnetic panels show all cells satisfying $|E|\geq5\langle|E|\rangle$ and $|B|\geq5\langle|B|\rangle$, respectively, with averages over the full box at each measurement time.
These spatial concentrations are consistent with localized defects, whose magnetic and electric charges are determined by the flux tests described below.

We next apply independent magnetic and electric flux tests to determine which localized objects are dyons.
We identify magnetic candidates as elementary cubes with nonzero integer magnetic charge $n_c$ defined in Eq.~(\ref{eq:magnetic-integer}).
The values $n_c = \pm 1$ correspond to magnetic charges $Q_m = \pm 4 \pi /g$, whereas $n_c = 0$ denotes zero net magnetic charge in the cube. 
Electric charge is measured separately as the flux through nested closed cubic surfaces.
The electric observable is constructed from adjacent time slices and includes the change in the Higgs direction.
The reference Higgs magnitude is $v=2$, and projected-field measurements require $|\bm\phi|>0.001v=0.002$.

A dyon label requires a persistent electric signal around a magnetically qualified candidate.
At each measurement time, we track the candidate backward through five samples spanning a conformal-time interval of approximately $1.25$.
Centering the measurement surfaces on a magnetically qualified elementary cube, we test the electric-flux profile at physical half-widths $\{2,3,4\}r_{\mathrm{core}}$ with $r_{\mathrm{core}}=1/\sqrt{2}$. 
The cuts require an amplitude above a numerical selection threshold $Q_{\min}=0.05(4\pi/g)\simeq0.6283$, a plateau tolerance of $0.25$, background significance of at least $3$, and relative temporal scatter no larger than $0.5$.
At least four of the five samples must pass these cuts with the same electric sign.

\begin{figure*}[!tp]
 \centering
 \includegraphics[width=0.94\textwidth,height=0.76\textheight,keepaspectratio]{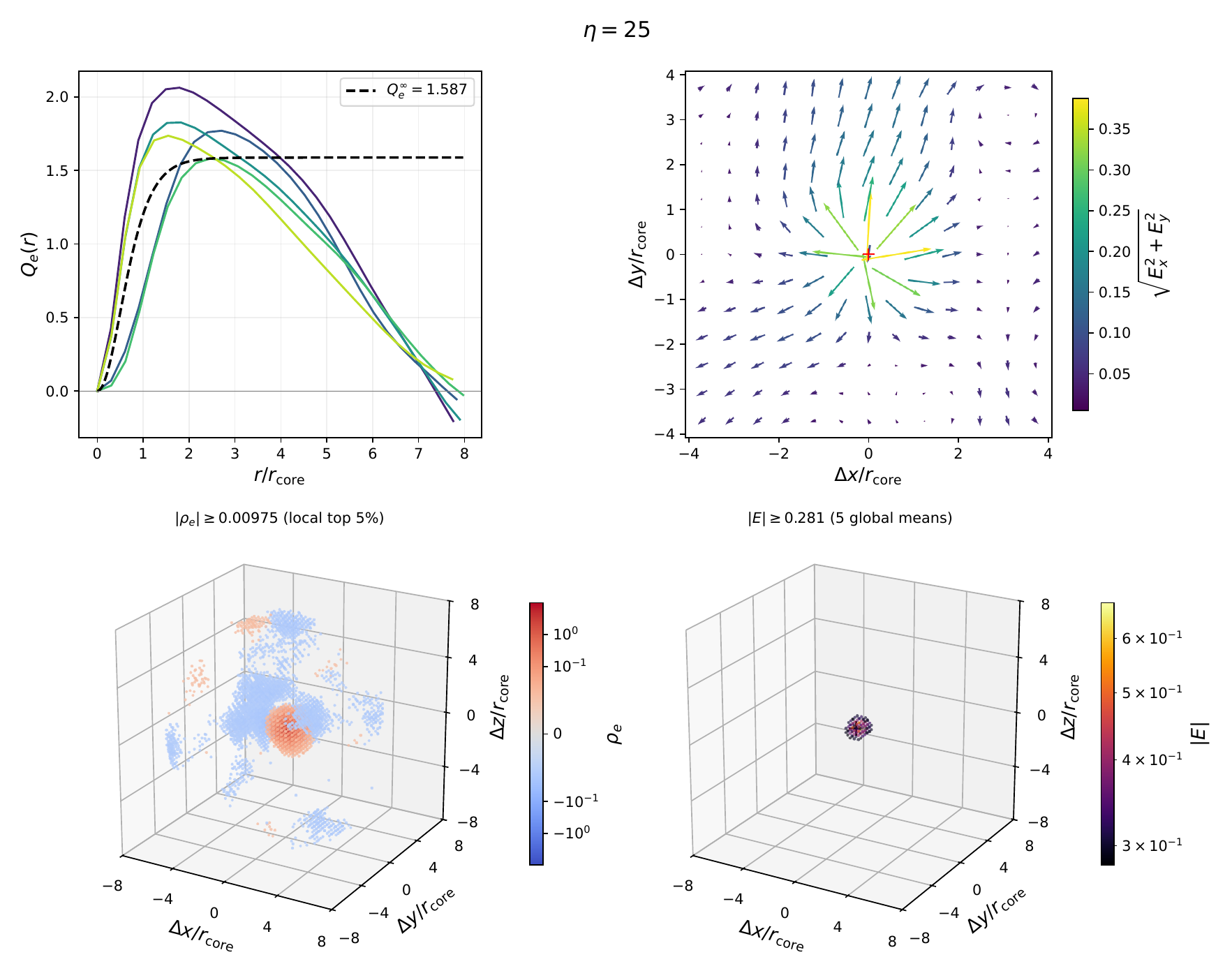}
\caption{Local fields of a dyon at $\eta=25$, whose charge evolution is shown in Fig.~\ref{fig:profiles40}.
The upper left panel shows five window charge profiles from the earliest sample (purple) to the endpoint (yellow), together with the static Julia--Zee cubic flux (black dashed) at $Q_e^\infty=\widehat Q_e=+1.587$.
Here $r$ is the physical half-width of the cubic flux surface centered on the candidate's elementary cube, shown in units of $r_{\mathrm{core}}$.
The static solution uses $g=1$, $\lambda=0.25$, and $v=2$, with no additional amplitude or length-scale fit.
The upper right panel shows $(E_x,E_y)$ in the $xy$ plane through the cube center, marked by a red cross.
Coordinates are physical offsets in units of $r_{\mathrm{core}}$.
The lower left panel shows the $5\%$ of points with the largest electric charge-density magnitudes $|\rho_e|$ in the local sphere, colored by the signed charge density.
The lower right panel shows physical electric-field magnitude above five times the full-box mean.
Both lower panels show points within a sphere of physical radius $8r_{\mathrm{core}}$, centered on the black cross.}
 \label{fig:local26}
\end{figure*}

\begin{figure*}[!tp]
 \centering
 \includegraphics[width=0.94\textwidth,height=0.76\textheight,keepaspectratio]{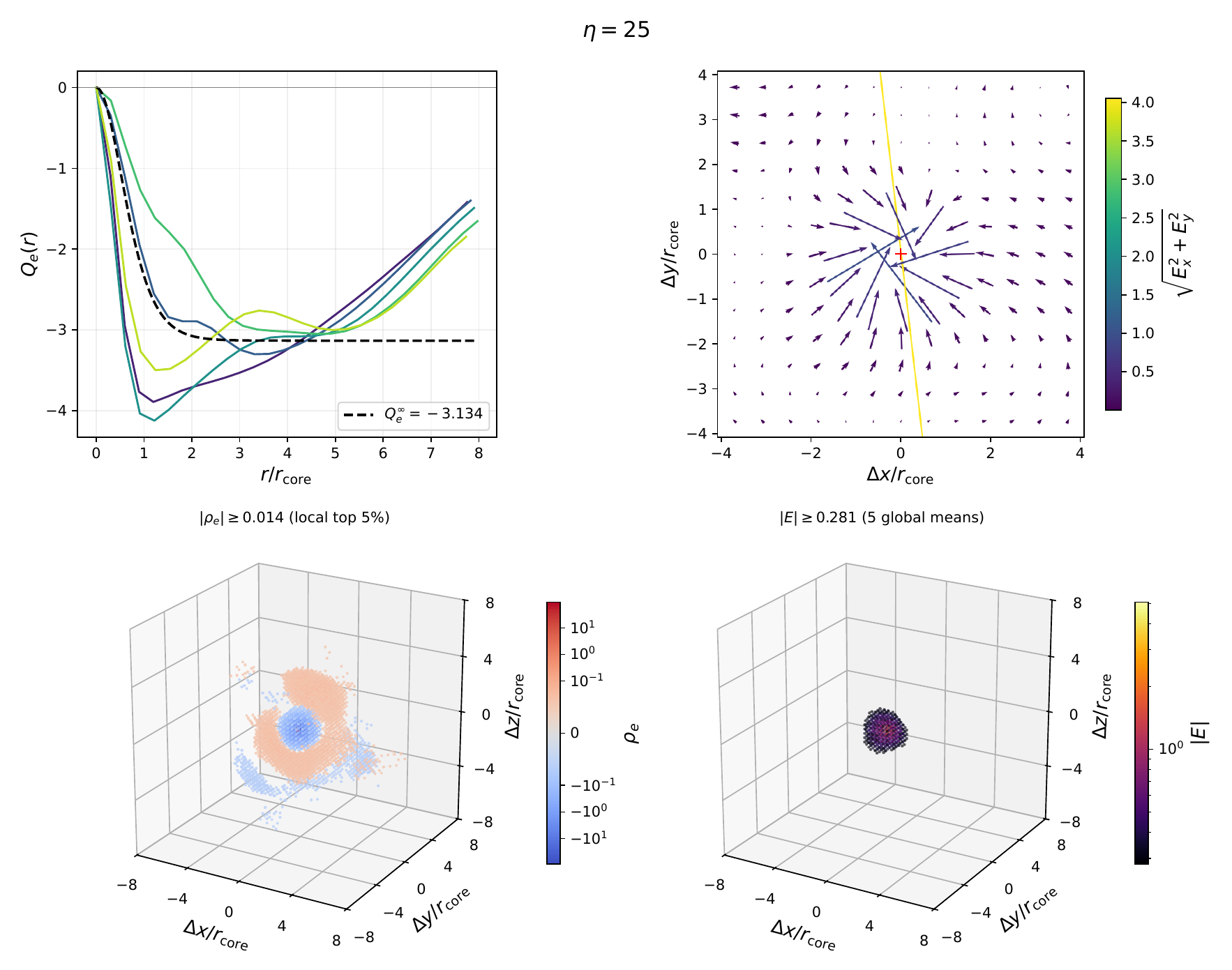}
\caption{Local fields of a second dyon at $\eta=25$, with the same conventions as Fig.~\ref{fig:local26}.
The charge history of this same dyon is shown in Fig.~\ref{fig:profiles25}.
The static comparison uses $Q_e^\infty=\widehat Q_e=-3.134$.
}
 \label{fig:local16}
\end{figure*}

Figures~\ref{fig:local26} and \ref{fig:local16} show two identified dyons at $\eta=25$, with each candidate's magnetically qualified elementary cube centered at the origin. 
The solid curves in the upper left panels show the electric charge enclosed by cubes of physical half-width $r$ at five sample times. 
Although the profiles fluctuate between samples, the enclosed electric charge remains nonzero at $r = \mathcal{O}(1)$. 
Their electric charges, averaged over the measurement half-widths and valid samples within the time window as defined in App.~\ref{app:charge}, are $\widehat Q_e=+1.587$ and $-3.134$, respectively.
The black dashed curves represent Julia--Zee solutions at $g=1$, $\lambda=0.25$, and $v=2$, where we set the static asymptotic charge to $Q_e(\infty)=\widehat Q_e$ and integrate its radial electric field over centered cubes with physical half-width $r$.

The upper right panels show two-dimensional slices of the electric field around the candidate dyons. 
Near the centers, the planar electric field is predominantly outward for the dyon in the upper-right panel of Fig.~\ref{fig:local26} and inward for that in Fig.~\ref{fig:local16}, consistent with their measured charge signs.

The bottom left panels show the $5\%$ of points with the largest charge-density magnitudes $|\rho_e|$ within a sphere of radius $8r_{\mathrm{core}}$, with red for positive and blue for negative charge. 
The bottom right panels show $|E|\geq5\langle|E|\rangle$. 
In both examples, the selected strong electric field is concentrated within $2r_{\mathrm{core}}$ of the center.
Moreover, the charge-density plots in the bottom left panels show compact concentrations of the corresponding sign near the cores, embedded in non-spherical structures containing both signs.
These observations connect the persistent electric flux to a localized charged configuration.


\begin{figure*}[!tp]
 \centering
 \includegraphics[width=0.98\textwidth,height=0.79\textheight,keepaspectratio]{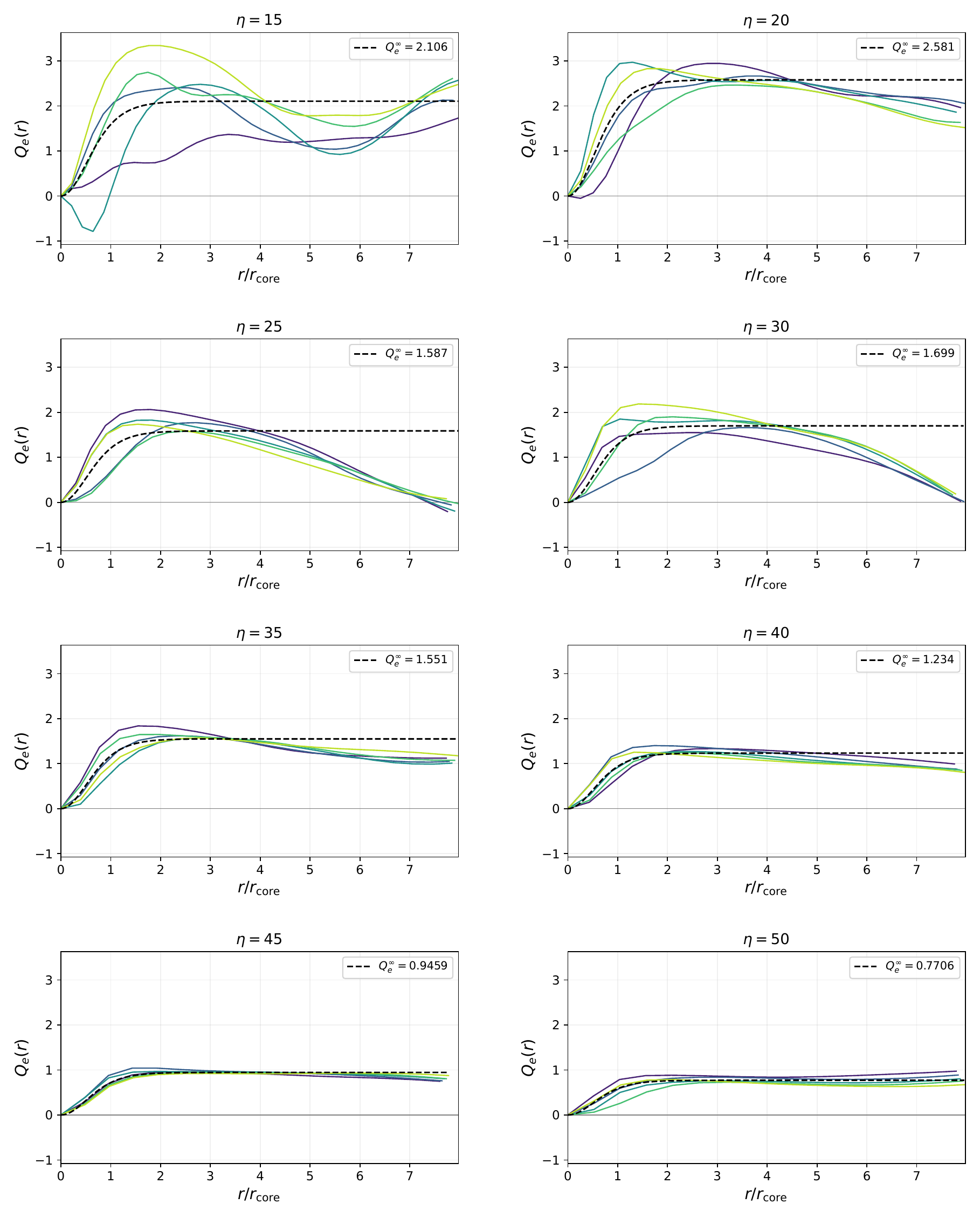}
\caption{Electric-charge evolution of the same dyon shown in Fig.~\ref{fig:local26}, in the $N=640$ run.
Panels progress from $\eta=15$ to $50$, from left to right and then downward.
The horizontal axis is the physical cube half-width in units of $r_{\mathrm{core}}$, and the vertical axis is enclosed electric charge.
Five simulation curves run from the earliest window sample (purple) to the endpoint (yellow).
Black dashed curves show static Julia--Zee cubic fluxes with the annotated $Q_e^\infty=\widehat Q_e$, at $g=1$, $\lambda=0.25$, and $v=2$.
Axes are shared across panels.
It remains classified as a dyon at all displayed times.}
 \label{fig:profiles40}
\end{figure*}
\begin{figure*}[!tp]
 \centering
 \includegraphics[width=0.98\textwidth,height=0.79\textheight,keepaspectratio]{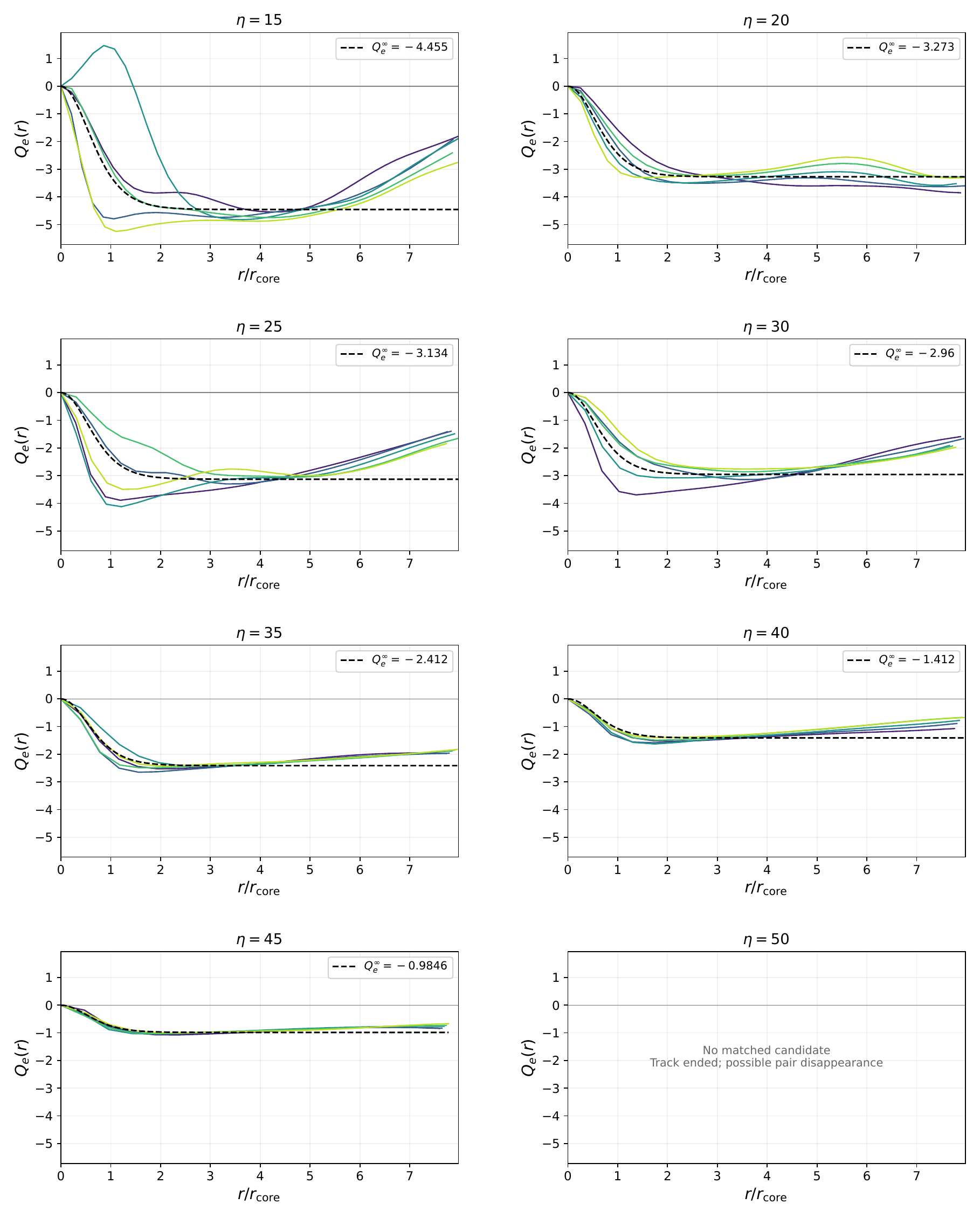}
\caption{Electric-charge evolution of the same dyon shown in Fig.~\ref{fig:local16}, with the same plotting conventions as Fig.~\ref{fig:profiles40}.
The final panel is empty because the track has no matched candidate at $\eta = 50$.}
 \label{fig:profiles25}
\end{figure*}

The opposite-sign charge surrounding each core can reduce the enclosed electric flux as the measurement surface grows.
This is visible in the departure from the isolated static profiles at larger cube half-widths in the upper left panels.
The surrounding charge also provides a possible source for discharge through absorption.
Next, we turn to the charge histories to determine whether the electric-charge magnitude actually decreases with time.

To associate objects between snapshots at different times, we minimize the total squared periodic comoving displacement among candidates of the same magnetic sign under a one-to-one assignment.
When matching candidates between snapshots, we allow a maximum comoving separation of $0.2$ times their conformal-time interval.

Figures~\ref{fig:profiles40} and \ref{fig:profiles25} follow the dyons shown in Figs.~\ref{fig:local26} and \ref{fig:local16}, respectively, starting at $\eta=15$. 
Their electric-charge magnitudes decrease at late times.
The dyon in Fig.~\ref{fig:local26} remains classified as a dyon through the final snapshot, with $\widehat Q_e=+1.587$, $+1.234$, and $+0.7706$ at $\eta = 25$, $40$, and $50$.
The decrease need not be monotonic between individual measurements.
For the dyon in Fig.~\ref{fig:local16}, the measured charge changes from $-3.134$ at $25$ to $-1.412$ at $40$ and $-0.9846$ at its last saved detection at $45$.
No matched candidate remains at $50$.
We find that an opposite-magnetic-charge neighbor disappears in the same interval, making this a possible pair-annihilation event.

As a numerical check, App.~\ref{app:static-charge-retention} follows dyons initialized from static profiles without added ambient charged excitations.
Their enclosed electric charges remain close to the initial values even when the physical lattice spacing exceeds the reference core size.
This check suggests that the charge decrease shown in Figs.~\ref{fig:profiles40} and \ref{fig:profiles25} is not simply an artifact of the increasing physical lattice spacing.

\subsection{Evolution of the electric-charge distribution}
\label{subsec:charge-distribution}

We next examine how the individual charge histories are reflected in the evolution of the dyon population and its electric-charge distribution.
Table~\ref{tab:counts640} lists the monopole and dyon counts at eight output times for the $N=640$ run.
$N_{\mathrm{mag}}$ counts all magnetic candidates, whereas $N_{\mathrm{qual}}$ counts those that pass the magnetic isolation and flux tests at the endpoint of each time window and are selected for electric-charge analysis.
$N_{\mathrm{dyon}}$ counts those that satisfy the dyon criteria.
The raw magnetic count decreases from $44$ at $\eta=15$ to $28$ at $50$.
The dyon count initially rises from $18$ to $20$ at $20$ and then falls to $1$ at $50$.
The dyon fractions of the raw magnetic population are $19/34\simeq0.559$, $8/30\simeq0.267$, and $1/28\simeq0.0357$ at $25$, $40$, and $50$.
The reduction in electrically selected objects therefore does not require the disappearance of most magnetic defects.

\begin{table}[t]
\caption{Counts for the $N=640$ realization.
$N_+$ and $N_-$ count the two magnetic signs, $N_{\mathrm{mag}}=N_++N_-$, and $N_{\mathrm{qual}}$ denotes magnetically qualified endpoints eligible for time-window analysis.
}
 \label{tab:counts640}
 \centering
 \begin{tabular}{@{}rrrrrr@{}}
  \toprule
  $\eta$ & $N_+$ & $N_-$ & $N_{\mathrm{mag}}$ & $N_{\mathrm{qual}}$ & $N_{\mathrm{dyon}}$ \\
  \midrule
  $15$ & $22$ & $22$ & $44$ & $32$ & $18$ \\
  $20$ & $17$ & $17$ & $34$ & $32$ & $20$ \\
  $25$ & $17$ & $17$ & $34$ & $32$ & $19$ \\
  $30$ & $16$ & $16$ & $32$ & $32$ & $16$ \\
  $35$ & $16$ & $16$ & $32$ & $30$ & $12$ \\
  $40$ & $15$ & $15$ & $30$ & $30$ & $8$ \\
  $45$ & $14$ & $14$ & $28$ & $28$ & $3$ \\
  $50$ & $14$ & $14$ & $28$ & $26$ & $1$ \\
  \bottomrule
 \end{tabular}
\end{table}

\begin{figure}[t]
 \centering
 \includegraphics[width=0.48\textwidth]{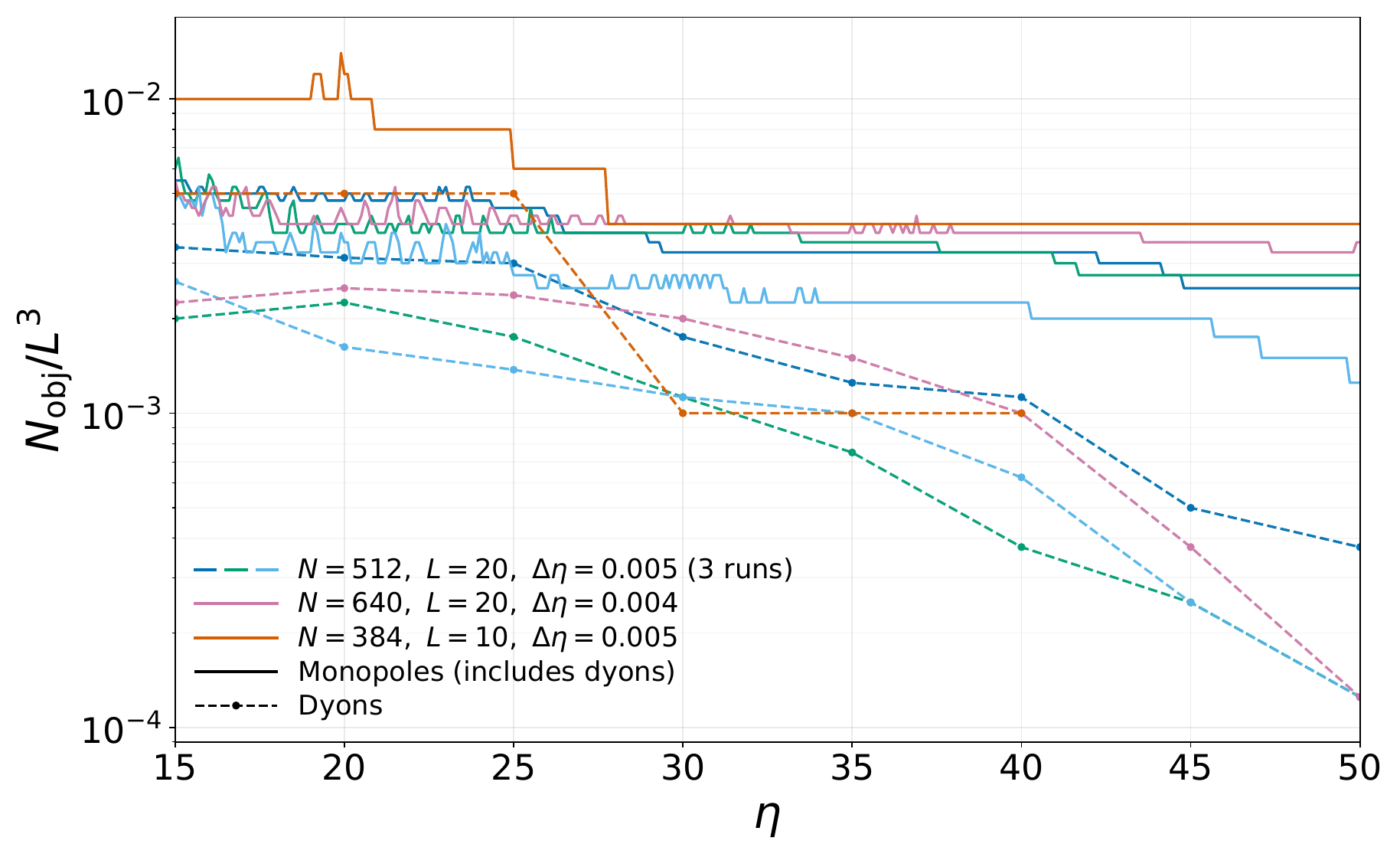}
\caption{Comoving monopole and dyon number densities for five runs.
Dashed curves show the dyon counts ($N_{\mathrm{obj}}=N_{\mathrm{dyon}}$) divided by each run's comoving volume $L^3$.
Solid curves show magnetic candidates ($N_{\mathrm{obj}}=N_{\mathrm{mag}}$), including dyons.
The legend identifies the lattice size, box size, and integration time step.
The three colors grouped under $N=512$ distinguish independent initial realizations.
Raw counts are sampled at intervals of $0.1$ in conformal time, while dyon classification is performed only at the endpoints indicated by the dots.
The zero dyon counts of the $N=384$ run at $45$ and $50$ are omitted from the logarithmic axis.}
 \label{fig:density-comparison}
\end{figure}

In addition to the above $N=640$ run, we perform three $N=512$, $L=20$, $\Delta\eta=0.005$ runs with different seeds and one $N=384$, $L=10$, $\Delta\eta=0.005$ run. 
Figure~\ref{fig:density-comparison} shows the comoving densities $N_{\mathrm{mag}}/L^3$ and $N_{\mathrm{dyon}}/L^3$ separately for each run, including both magnetic signs and counting dyons within the magnetic population.
The dyon density decreases more strongly at late times, while a substantial magnetic population remains.
The curves are not an ensemble average, and their fluctuations reflect the different volumes, resolutions, and initial realizations.

\begin{figure*}[!tp]
 \centering
 \includegraphics[width=0.94\textwidth,height=0.75\textheight,keepaspectratio]{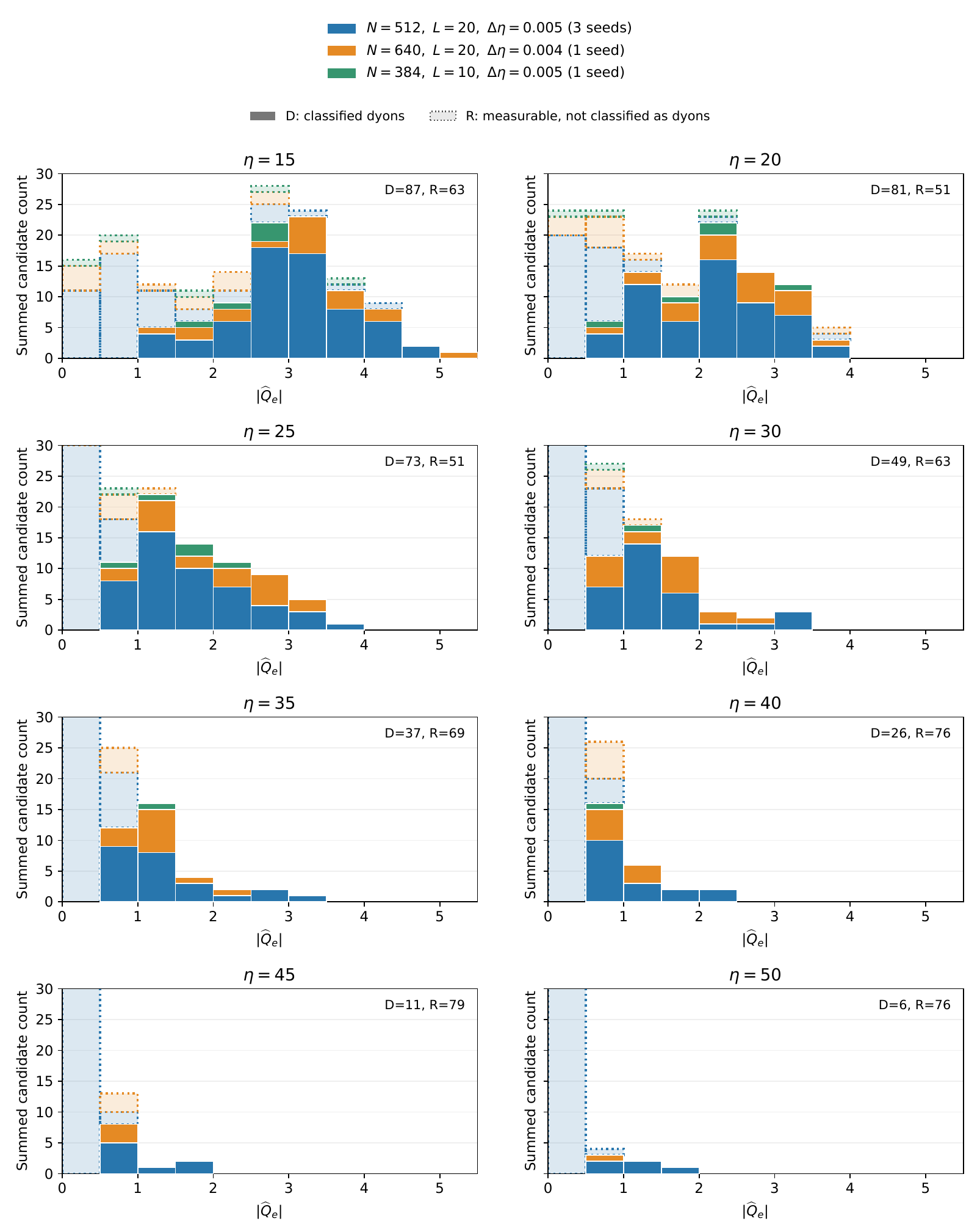}
\caption{
Stacked histograms of $|\widehat Q_e|$ for magnetic candidates with measurable electric charge, summed over the five runs described in Sec.~\ref{subsec:simulation-setup}.
Solid fills show identified dyons, and pale fills with dotted outlines show candidates that fail the dyon selection criteria.
The panel annotations $\mathrm D$ and $\mathrm R$ give their respective total counts, including contributions above the displayed vertical range.
Blue combines the three $N=512$ runs with different seeds, while orange and green show the $N=640$ and $N=384$ runs, respectively.
Some low-charge bins exceed the vertical limit because many candidates have charges below the amplitude or background-significance thresholds.
Bins have width $0.5$ and show raw counts, without volume or sample-size normalization.
Candidates without a valid charge classification are excluded.
}
 \label{fig:charge-histograms}
\end{figure*}

We then turn from the number densities to the electric-charge distribution. 
Figure~\ref{fig:charge-histograms} shows stacked histograms of magnetic candidates with measurable electric charge from the five runs in bins of $|\widehat Q_e|$ at each time.
Solid fills show identified dyons, while pale fills with dotted outlines show candidates whose charges were measured but which failed the dyon selection criteria in App.~\ref{app:charge}.
Colors distinguish the simulation parameter sets, with counts from runs that differ only in the random seed combined in the same color.
The bin width is $0.5$, and positive and negative charges contribute to the same magnitude distribution.
The large populations at small $|\widehat Q_e|$ exceed the plotted vertical range in several panels.
These candidates have electric signals that are too small in absolute magnitude or relative to the surrounding fluctuations to be distinguished reliably from neutral monopoles by our selection.

The late-time histograms contain fewer identified dyons and extend over a smaller observed dyon charge range.
The selected dyon counts are $87,81,73,49,37,26,11,6$ at $\eta=15,20,25,30,35,40,45,50$, respectively.
These histograms show raw counts summed over the five runs, without normalization by volume or sample size.
Although they do not give a probability density or an equally weighted average over runs, the combined distributions show a qualitative trend toward smaller electric charges and fewer identified dyons at late times.

\section{Discussion and conclusions}
\label{sec:conclusions}

We have demonstrated the formation of Julia--Zee dyons in classical lattice simulations of an $\mathrm{SU}(2)\to\mathrm{U}(1)$ theory in an expanding background.
Starting from thermally prepared gauge and Higgs fields with ultraviolet filtering, we followed symmetry breaking and the subsequent electric-charge evolution of the magnetic defects.
The earlier charge distributions extend to several units of $g$, including charges exceeding $4g$, while the observed charge range and the number of identified dyons decrease at late times.
In the $N^3=640^3$ run, the dyon count falls from $20$ at $\eta=20$ to $1$ at $50$, whereas the magnetic count decreases from $34$ to $28$ over the same interval.
Electric-charge relaxation thus accompanies the persistence of a substantial magnetic population, although these results do not establish that the cosmological monopole abundance is unchanged by dyon formation.

The tracked charge profiles and spatial field distributions are consistent with charge exchange with the surrounding fields.
Opposite-sign charge near a core can contribute to discharge, and we also find possible pair-annihilation events.

Our simulations establish the qualitative behavior of dyon formation and subsequent discharge, but do not provide a quantitative prediction for the surviving monopole abundance or the time dependence of the dyon population.
Studies of monopole relic abundances illustrate the importance of the thermal history and monopole annihilation rates~\cite{Zeldovich:1978wj,Preskill:1979zi,Martins:2008ks,Liveoak:2026muy,Liveoak:2026dtx}.
Recent studies also constrain reheating using light monopoles~\cite{Pesut:2026fyl} and propose enhanced annihilation through a time-dependent gauge kinetic term~\cite{Perri:2025vtn}.
The present simulations nevertheless demonstrate that symmetry breaking produces identifiable dyons that persist for a finite period and subsequently lose electric charge.
The charge histories and surrounding opposite-sign charge distributions support a picture in which the dyons discharge by absorbing charge from their environment.
Establishing this behavior in dynamical classical simulations provides a basis for future quantitative studies of dyon production and relaxation.

Applying this result to a quantum theory requires care.
At vanishing vacuum angle, the bosonic adjoint theory has discrete dyon charge eigenvalues $Q_e=g n_e$, whereas our classical evolution allows continuous electric charge.
Treating the charge spacing as a small correction would require $|Q_e|/g\gg1$, and the observed charges of only several units do not provide that hierarchy.
The calculation therefore motivates quantum dyon production during symmetry breaking but does not give a controlled prediction for the populations of integer charge states.
Nor does the classical decrease in charge establish that only neutral monopoles survive in the quantum theory.
That outcome depends on the rates of charge exchange between the allowed states and their competition with expansion.
Light charged fermions can also alter dyon charge profiles and stability and allow charge exchange and fermion emission~\cite{Blaer:1981ui,Callan:1982ah,Callan:1982au,Marciano:1983pu,Hook:2024vhf}, and their effects are absent from our purely bosonic simulations.

These questions are relevant to axion cosmology through the Witten effect~\cite{Fischler:1983sc,Kawasaki:2015lpf,Jeong:2023gjc,Co:2025lcs}.
If the vacuum angle $\theta$ is promoted to an axion-dependent field, the electric charge depends on both $\theta$ and the integer $n_e$ specifying the electric charge state.
The axion-dependent energy and force therefore depend on which charge states are populated and how rapidly transitions between them occur.
For an approximately quadratic charge energy and fixed populations, changing the distribution of $n_e$ can shift the preferred axion value without changing the curvature at fixed magnetic abundance.
Calculations of axion dissipation through transitions between dyon levels with light charged fermions~\cite{Co:2025lcs} illustrate why both the charge spectrum and the transition rates matter.

Recently, it has been discussed that 
an asymmetric population of higher-charge dyons at fixed magnetic charge can also enhance the interaction between monopole dark matter and axion dark energy~\cite{Bagherian:2026jxz}.
The formation and subsequent charge relaxation observed in our classical simulations highlight the need to establish whether such populations survive charge exchange with their surroundings.
Determining the resulting axion evolution and relic abundance will require an extension that includes axion dynamics and transitions between quantized charge states.

\section*{Acknowledgements}
This work was supported by JSPS KAKENHI Grant Number 23K13092.
Some of the numerical computations in this work were performed using the supercomputing resources provided by the Cyberscience Center, Tohoku University.

\appendix
\section{Numerical methods}
\label{app:preparation}

In this appendix, we explain the detailed numerical algorithms for our $(3+1)$-dimensional classical lattice simulations. 

\subsection{Lattice variables and thermal preparation}

For the dynamical gauge field, we use the link formulation introduced by Wilson and its Hamiltonian form developed by Kogut and Susskind~\cite{Wilson:1974sk,Kogut:1974ag}.

Here $x$ labels a site of a periodic $N^3$ lattice, $i=1,2,3$ labels a spatial direction, and $\hat i$ denotes one lattice step in that direction.
The comoving spacing is $\Delta x=L/N$, where $L$ is the box side length, and the physical spacing is $R\Delta x$, where $R$ is the scale factor.

At each site, $\bm\phi_x$ and $\bm\Pi_x$ are the three adjoint components of the Higgs field and its lattice canonical momentum.
The index $a=1,2,3$ labels these internal components, and repeated internal indices are summed.

A link $U_{x,i}$ transports a field at $x+\hat i$ to the gauge frame at $x$.
The conjugate link momentum $\bm P_{x,i}$ transforms as an adjoint vector at $x$.
For any adjoint vector $\bm X$, the transport operation is defined by
\begin{align}
 (\operatorname{Ad}_U\bm X)^a\sigma^a
 &=U(X^a\sigma^a)U^\dagger .
\end{align}

A plaquette is the ordered product of four links around an elementary lattice square.
We use the Wilson plaquette action for the gauge field and transported nearest-neighbor differences for scalar gradients.
The symbols $\bm E$ and $\bm B$ denote physical projected fields, defined in App.~\ref{app:charge}.

We write the link as $U_{x,i}=u_0\mathbf{1}+\ii u_a\sigma^a$, where $u_0$ and $u_a$ are real, $\sigma^a$ are the Pauli matrices, and $u_0^2+\bm u^2=1$.
To begin the thermal preparation, we initialize $U_{x,i}=\mathbf{1}$ and draw each scalar component from a zero-mean Gaussian of width $0.08$.
The small random initial scalar values provide spatial fluctuations without selecting a preferred direction in internal space.
They serve only as a starting configuration for the thermal preparation and do not themselves represent a thermal equilibrium state.

To prepare thermal fluctuations, we repeatedly refresh the momenta and then evolve the fields and momenta along a Hamiltonian trajectory.
At each refresh, we retain the gauge and scalar fields, redraw their conjugate momenta from the Gaussian distributions with variances set by $T$, 
\begin{align}
 \left\langle(P_{x,i}^{a})^2\right\rangle
    =\frac{T\Delta x}{g^2},\qquad
 \left\langle(\Pi_x^a)^2\right\rangle
    =T(\Delta x)^3 R^2, 
 \label{eq:refresh-variances}
\end{align}
and enforce the Gauss constraint.
We then evolve both the fields and momenta for a fixed trajectory duration, allowing the momentum fluctuations to generate fluctuations in the field configurations through their coupled dynamics.
The final field configuration of each trajectory becomes the starting configuration for the next refresh.
Thus the $40$ trajectories form a single preparation sequence rather than $40$ independent initial configurations.

This algorithm corresponds to the proposal stage of hybrid Monte Carlo~\cite{Duane:1987de}.
We do not apply its Metropolis acceptance test, which corrects finite-step integration errors in equilibrium sampling.
Thus the $40$ trajectories constitute a finite preparation of thermal fluctuations.

\subsection{Enforcing the Gauss constraint}

Random momentum draws generally violate Gauss's law.
In the above canonical normalization, its residual, which vanishes when the constraint is satisfied, is
\begin{align}
 \bm G_x={}&\sum_i\left[\bm P_{x,i}
 -\operatorname{Ad}_{U_{x-\hat i,i}^{\dagger}}
                    \bm P_{x-\hat i,i}\right]
 -\bm\phi_x\mathbin{\times}\bm\Pi_x .
 \label{eq:lattice-gauss}
\end{align}
The cross product acts on the three internal components.
At fixed $U$ and $\phi$, let us collect all link and scalar momenta into one vector $p=(P,\Pi)$ and all residuals into $G$.
Eq.~(\ref{eq:lattice-gauss}) then defines the linear map $C$ through $G=Cp$.
We define the diagonal matrix $W$ through the unconstrained momentum covariance $TW$, with link and scalar entries given by
\begin{align}
 W_P=\frac{\Delta x}{g^2},\qquad
 W_\Pi=(\Delta x)^3R^2, 
\end{align}
from Eq.~(\ref{eq:refresh-variances}).
We impose the Gauss constraint while minimizing the change in the momenta relative to their thermal fluctuation scales.
Specifically, we minimize $\delta p^{\mathsf T}W^{-1}\delta p$ subject to $C(p+\delta p)=0$.
Introducing one multiplier $\omega_x^a$ for each constraint gives 
\begin{align}
 (CWC^{\mathsf T})\omega=G,\qquad
 p\longleftarrow p-WC^{\mathsf T}\omega .
 \label{eq:gauss-projection}
\end{align}
Here $\mathsf T$ denotes transpose, and $\omega$ collects the multipliers.
Substitution into $Cp$ shows that the corrected momenta satisfy the constraint, $G = Cp = 0$, to the linear-solver tolerance.
The coordinates $U$ and $\phi$ remain fixed.

We solve this linear system with the conjugate-gradient method~\cite{Hestenes:1952mok}.
To accelerate convergence, we rescale the residual components by the inverse diagonal of $CWC^{\mathsf T}$.
For component $a$ at site $x$, this diagonal is $6W_P+W_\Pi[\bm\phi_x^2-(\phi_x^a)^2]$.
A small positive numerical floor prevents division by a vanishing diagonal entry.
We recompute the residual during the solve and check Eq.~(\ref{eq:lattice-gauss}) directly after correcting the momenta.

After the UV filtering described below, we require the root-mean-square (RMS) residual to satisfy
\begin{align}
 G_{\mathrm{rms}}^{\mathrm{final}}\leq
 \max\left(2\times10^{-5}G_{\mathrm{rms}}^{\mathrm{initial}},
                 2\times10^{-5}\right).
 \label{eq:gauss-tolerance}
\end{align}
Here $G_{\mathrm{rms}}=[\sum_{x,a}(G_x^a)^2/(3N^3)]^{1/2}$, and ``initial'' and ``final'' denote the Gauss residuals immediately before and after a single momentum projection, respectively.

\subsection{Gauge-covariant UV filtering}
\label{app:evolution}

The UV filtering stage is applied once, after the thermal preparation.
It smooths short-wavelength fluctuations by a dissipative flow, separate from physical time evolution.
The link smoothing uses the Wilson-action gradient underlying Wilson flow~\cite{Luscher:2010iy}.

We smooth the fields by repeating a filtering step, which we call a sweep, over the entire lattice.
Each sweep updates the links, scalar field, and momenta with a dimensionless smoothing step size $\epsilon_{\mathrm f}$.
We first compute the Wilson gauge force, obtained from the variation of the Wilson plaquette action, using the link configuration at the start of the sweep.
We then update each link by multiplication with an $\mathrm{SU}(2)$ exponential constructed from this force.
After this update, we normalize its four real components so that $u_0^2+\bm u^2=1$, correcting numerical roundoff errors.

Using the updated links, we then smooth the scalar field $\phi$, its momentum $\Pi$, and each of the three link-momentum fields $P_i$.
For any of these adjoint fields $X$, we apply the diffusion update
\begin{align}
 (\mathcal L_U X)_x
 &=\sum_i\left[
 \operatorname{Ad}_{U_{x,i}}X_{x+\hat i}
 +\operatorname{Ad}_{U_{x-\hat i,i}^{\dagger}}X_{x-\hat i}
 -2X_x\right],\\
 X_x&\longleftarrow X_x+\epsilon_{\mathrm f}(\mathcal L_U X)_x.
\end{align}
Here, $\mathcal L_U$ is the dimensionless gauge-covariant lattice Laplacian, and $\operatorname{Ad}_U$ denotes the adjoint action of the link matrix.
The link matrices transport the six neighboring field values to the gauge frame at $x$, allowing them to be compared with $X_x$.
The update reduces local variations while preserving the gauge transformation properties of the fields.
All right-hand sides for a given field are evaluated before updating that field, and the three $P_i$ fields are updated simultaneously.

The input cutoff specifies a smooth filter through the total dimensionless flow time
\begin{align}
 \tau_{\mathrm f}
 =\frac{1}{2(\Delta x R_{\mathrm{th}} k_{\mathrm{cut}})^2}.
\end{align}
Here $R_{\mathrm{th}}=1$ is the scale factor during preparation, and $k_{\mathrm{cut}}$ is a physical wave number.
In the weak-field, long-wavelength limit, the diffusion equation suppresses a mode of physical wave number $k$ by $\exp[-\tau_{\mathrm f}(R_{\mathrm{th}}\Delta x k)^2]$.
Thus $k_{\mathrm{cut}}$ sets a smooth suppression scale.
For $k_{\mathrm{cut}}=8$ and $\Delta x=20/640$, $\tau_{\mathrm f}=8$.
We use $\tau_{\mathrm f}/\epsilon_{\mathrm f} = 160$ sweeps with $\epsilon_{\mathrm f}=0.05$.
Filtering all momentum channels can change the Gauss constraint, so Eq.~(\ref{eq:gauss-projection}) is applied after the final sweep.

\subsection{Time integration and the mass quench}

We evolve the fields during and after the quench using the velocity-Verlet algorithm adapted to $\mathrm{SU}(2)$ link variables. 
It consists of a half-step momentum update, a full-step coordinate update, and a second half-step momentum update.
It has second-order accuracy in time. 
Let $\eta_n$ denote the conformal time at step $n$, with $\eta_{n+1}=\eta_n+\Delta\eta$.
The forces follow by varying the lattice Hamiltonian with respect to the scalar field and link rotations.
The first half-step evaluates them at the old fields, mass, and scale factor.
After updating the momenta to the half step, we update the links and scalar field as
\begin{align}
 U_{x,i}^{n+1}
 &=\exp\!\left[\frac{\ii\Delta\eta\,g^2}{2\Delta x}
                  P_{x,i}^{a,n+1/2}\sigma^a\right]U_{x,i}^{n},
 \label{eq:link-drift}\\
 \bm\phi_x^{n+1}
 &=\bm\phi_x^n+
 \frac{\bm\Pi_x^{n+1/2}}{(\Delta x)^3}
 \int_{\eta_n}^{\eta_{n+1}}\frac{d\eta}{R^2(\eta)}.
 \label{eq:scalar-drift}
\end{align}
The superscript $n+1/2$ labels the momenta after the first half-step.
The second half-step evaluates the forces at the new coordinates and at time $\eta_{n+1}$.
Link quaternions are normalized after the drift to remove accumulated floating-point norm errors.
In the force evaluation, the Wilson, scalar-gradient, and scalar-potential terms carry the lattice prefactors $4/(g^2\Delta x)$, $R^2\Delta x$, and $R^4(\Delta x)^3$, respectively.

We denote the conformal time measured from the onset of the quench by $\eta_{\mathrm{quench}}$. 
During the quench, we take $R=R_0(1+h_0\eta_{\mathrm{quench}})$, where $R_0=1$ is the initial scale factor and $h_0=0.2$ sets its slope in conformal time.
Accordingly, the scalar drift uses the exact integral
\begin{align}
 \int_{\eta_n}^{\eta_{n+1}}\frac{d\eta}{R^2(\eta)}
       =\frac{\Delta\eta}{R(\eta_n)R(\eta_{n+1})}.
\end{align}
During the quench we interpolate the mass as
\begin{align}
 m^2(\eta_{\mathrm{quench}})=m_i^2+\frac{m_f^2-m_i^2}{2}
 \left[1+\tanh\!\left(\frac{\eta_{\mathrm{quench}}-\eta_c}{w}\right)\right].
 \label{eq:mass-quench}
\end{align}
Here $m_i^2=1$ and $m_f^2=-1$ are the initial and final mass parameters.
We take $\eta_c=0.45\times5$ and $w=0.12\times5$, which set the center and characteristic duration of the transition, respectively.

After the quench, the shifted time coordinate $\eta=\eta_{\mathrm{quench}}-5$ gives $R=2+0.2\eta$.
The post-quench phase uses $m_f^2$ exactly and continues the same expansion history.

After the time evolution, we verify that $G_{\mathrm{rms}}<0.002$ and that the link-norm deviation from $u_0^2+\bm u^2=1$ is below $2\times10^{-4}$.

\section{Projected fields and magnetic and electric charges}
\label{app:charge}

In this appendix, we describe how we measure magnetic and electric charges and identify dyons on the lattice.
We first define the electric and magnetic fields associated with the residual $\mathrm{U}(1)$ gauge symmetry in the Higgs phase.

\subsection{Electric and magnetic fields of the residual U(1)}

We extract the residual $\mathrm{U}(1)$ field by projecting onto the local Higgs direction, following the lattice construction of Refs.~\cite{Davis:2000kv,Edwards:2009bw}.

At a site with nonzero Higgs magnitude, we define $\bm n \equiv \bm\phi/|\bm\phi|$.
We introduce a two-component complex vector $z$ satisfying $z^\dagger z=1$ and $z^\dagger\bm\sigma z=\bm n$.
The spatial projected link amplitude is
\begin{align}
 \ell_i(x)=z_x^\dagger U_{x,i}z_{x+\hat i}.
\end{align}
For an oriented spatial plaquette, its compact angle is
\begin{align}
 \theta_{ij}(x)=\operatorname{Arg}
 \left[\ell_i(x)\ell_j(x+\hat i)
       \ell_i^*(x+\hat j)\ell_j^*(x)\right],
 \label{eq:spatial-angle}
\end{align}
where $\operatorname{Arg}$ gives an angle in $(-\pi,\pi]$. 
The normalized eigenvector $z_x$ is defined only up to a local phase, $z_x\to e^{\mathrm{i}\alpha_x}z_x$.
These arbitrary phases cancel in the product of projected links around a closed loop.

In temporal gauge, the temporal $\mathrm{SU}(2)$ link is the identity, but the Higgs direction can change with time.
The projected temporal link is therefore $\ell_0(x)=z_x^\dagger(\eta_n)z_x(\eta_{n+1})$.
The temporal angle is
\begin{align}
 \theta_{0i}=\operatorname{Arg}
 \left[\ell_0(x)\ell_i(x,\eta_{n+1})
       \ell_0^*(x+\hat i)\ell_i^*(x,\eta_n)\right].
\end{align}
The physical projected fields are then given by 
\begin{align}
 E_i=\frac{2\theta_{0i}}{g\Delta x\Delta\eta R_{n+1/2}^2},
 \qquad B_i=\frac{\epsilon_{ijk} \theta_{jk}}{g(\Delta x)^2R_{n+1}^2},
 \label{eq:projected-fields}
\end{align}
where $\theta_{jk}=-\theta_{kj}$, and repeated spatial indices are summed.
Here $R_{n+1/2}$ is the scale factor at the temporal plaquette midpoint, and $R_{n+1}$ is its value on the final spatial slice.

\subsection{Magnetic charge and electric fluxes}

We identify magnetic charge from the outward sum of compact plaquette angles~\cite{DeGrand:1980eq}.
For a valid elementary cube $c$, define
\begin{align}
 n_c=\operatorname{nint}\!\left[
       \frac{1}{2\pi}\sum_{f\in\partial c}s_f\theta_f\right],
 \qquad Q_m(c)=\frac{4\pi}{g}n_c.
 \label{eq:magnetic-integer}
\end{align}
Here $\partial c$ denotes the six faces, $s_f=\pm1$ fixes their outward orientation, and $\operatorname{nint}$ rounds to the nearest integer.

Electric-charge profiles are direct outward flux sums through nested closed cubic surfaces centered on a candidate.
We denote by $C(r_e)$ a cube centered on the candidate with comoving half-width $r_e\Delta x$, where $r_e$ is a positive integer.
Its physical half-width is $r=r_e\Delta x R_{n+1/2}$, which is used as the horizontal coordinate in Figs.~\ref{fig:local26}--\ref{fig:profiles25}.
We denote the comoving electric field by $E_{\mathrm{com}}=R_{n+1/2}^2E$.
The enclosed electric charge is
\begin{align}
 Q_e(r)=(\Delta x)^2
          \sum_{\partial C(r_e)}s_f E_{\mathrm{com},f} .
 \label{eq:electric-flux}
\end{align}
Here $E_{\mathrm{com},f}$ is the electric component normal to a surface element, and $s_f$ fixes its outward sign.

We also determine the charge density by reconstructing the projected electric field from adjacent slices and evaluating its discrete divergence, 
\begin{align}
 \rho_e(x)=\frac{1}{R_{n+1/2}^3\Delta x}\sum_i
 \left[E_{i,\mathrm{com}}(x)-E_{i,\mathrm{com}}(x-\hat i)\right]. 
 \label{eq:local-electric-charge-density}
\end{align}
The discrete volume integral of this charge density equals the electric flux through the boundary.

\subsection{Dyon identification}

We require each magnetic candidate to be isolated from other magnetic charges.
The target physical half-widths are $r_j=(j+1)r_{\mathrm{core}}$, with $j=1,2,3$.
At each sample, we define the integer indices of the magnetic flux surfaces by
\begin{align}
 r_{m,j}
 &=\left\lceil\frac{r_j}{\Delta x R_{n+1}}+\frac12\right\rceil,
\end{align}
where $\lceil\cdot\rceil$ rounds upward to an integer.
The half-cell offset accounts for centering each surface on the elementary cube containing the candidate.
The isolation region contains all elementary cubes within $r_{m,3}$ lattice steps of the candidate in each coordinate direction, with distances evaluated using periodic boundary conditions.
All cubes in this region must have valid magnetic measurements, and only the central cube may carry nonzero magnetic charge.

At the endpoint of each time window, we additionally require valid magnetic flux measurements through all three surfaces.
Denoting these fluxes by $Q_{m,j}$ and the central cube charge by $Q_m(c)=4\pi n_c/g$, we require
\begin{align}
 |Q_{m,j}-Q_m(c)|
 \leq0.05\max\left(\frac{4\pi}{g},|Q_m(c)|,|Q_{m,j}|\right)
\end{align}
for every $j$.
A candidate passing these isolation and flux tests is counted in $N_{\mathrm{qual}}$.

Each magnetic candidate is tracked backward through five samples spanning approximately $1.25$ in conformal time.
The assignment first maximizes the number of matches and then minimizes their total squared displacement.

We identify a magnetic candidate as a dyon when it is accompanied by a persistent electric flux that exceeds surrounding fluctuations. 
We adopt the following dyon identification criteria. 
We denote by $Q_{e,kj}$ the candidate electric flux at sample $k$ and target half-width $r_j$, and by $\sigma_{kj}$ the corresponding background standard deviation, with $j=1,2,3$. 
Here, electric fluxes at $r_j$ are obtained by linear interpolation between adjacent native-grid half-widths, with both values required to be valid.
We estimate $\sigma_{kj}$ by linearly interpolating to $r_j$ the sample standard deviations of valid electric fluxes at native-grid half-widths, using up to $256$ randomly selected background centers with monopole-free isolation regions at all five samples.
For each magnetic candidate with all three finite fluxes, we define
\begin{align}
 Q_{e,k}&=\frac13\sum_j Q_{e,kj},\qquad
 b_k=\left(\frac13\sum_j\sigma_{kj}^2\right)^{1/2},\\
 p_k&=\frac{\left[\tfrac13\sum_j(Q_{e,kj}-Q_{e,k})^2\right]^{1/2}}
                {\max(|Q_{e,k}|,Q_{\min})},\qquad
 Q_{\min}=0.05\frac{4\pi}{g}.
\end{align}
Thus $Q_{e,k}$ is the mean enclosed charge over the three half-widths, $b_k$ is the RMS background fluctuation scale, and $p_k$ measures variation across these half-widths relative to the charge.
At a single sample, we require
\begin{align}
 p_k\leq0.25,\qquad |Q_{e,k}|\geq Q_{\min},\qquad
 \frac{|Q_{e,k}|}{b_k}\geq3.
 \label{eq:sample-cuts}
\end{align}
At least four of the five samples must satisfy the magnetic isolation tests and have valid flux measurements.
Otherwise the candidate is left unclassified.

We define $K$ to contain \emph{all} such eligible and measurable samples.
The reported charge is
\begin{align}
 \overline Q_{e,j}=\frac{1}{|K|}\sum_{k\in K}Q_{e,kj},\qquad
 \widehat Q_e=\frac13\sum_j\overline Q_{e,j} .
 \label{eq:reported-charge}
\end{align}
The symbol $|K|$ denotes the number of included samples, and $\overline Q_{e,j}$ is the time-averaged charge at half-width $r_j$.
We apply the same limits on variation across the half-widths and charge magnitude to the three time-averaged charges $\overline Q_{e,j}$ as to the charges measured at each individual sample.
For the averaged charge, we use the background scale $\bar b=[\sum_{k=1}^{5}\sum_j\sigma_{kj}^2/15]^{1/2}$ and require $|\widehat Q_e|/\bar b\geq3$.
To limit temporal variation, the sample standard deviation of $Q_{e,k}$ over $K$ must obey
\begin{align}
 \frac{\left[\sum_{k\in K}(Q_{e,k}-\widehat Q_e)^2/(|K|-1)\right]^{1/2}}
      {\max(|\widehat Q_e|,Q_{\min})}\leq0.5.
\end{align}
Finally, at least four samples must pass Eq.~(\ref{eq:sample-cuts}) and have the same nonzero sign as $\widehat Q_e$.
A dyon label is assigned only when all these conditions hold.

\section{Test of electric-charge retention on coarse lattices}
\label{app:static-charge-retention}

We test electric-charge retention by initializing a dyon-antidyon pair on a periodic three-dimensional lattice using the static Julia--Zee profiles of Sec.~\ref{sec:model}.
The target electric-charge magnitudes are $|Q_e|=4$.
The initial electric field is aligned with the magnetic field, normalized to the target charges, and projected onto the Gauss constraint.
We add no thermal or other nonequilibrium charged excitations to the surrounding fields and omit thermal preparation and the mass quench.
All runs use $N=256$, $g=1$, $\lambda=0.25$, $m^2=-1$, and $\Delta\eta=0.004$.
Evolution starts at the post-quench time $\eta=0$ with $R=2+0.2\eta$.
We vary $L=16,24,32,40,48$, so the physical lattice spacing is
\begin{align}
 a_{\mathrm{phys}}=R(\eta)\Delta x=(2+0.2\eta)\frac{L}{N}.
\end{align}

\begin{figure}[t]
 \centering
 \includegraphics[width=0.48\textwidth]{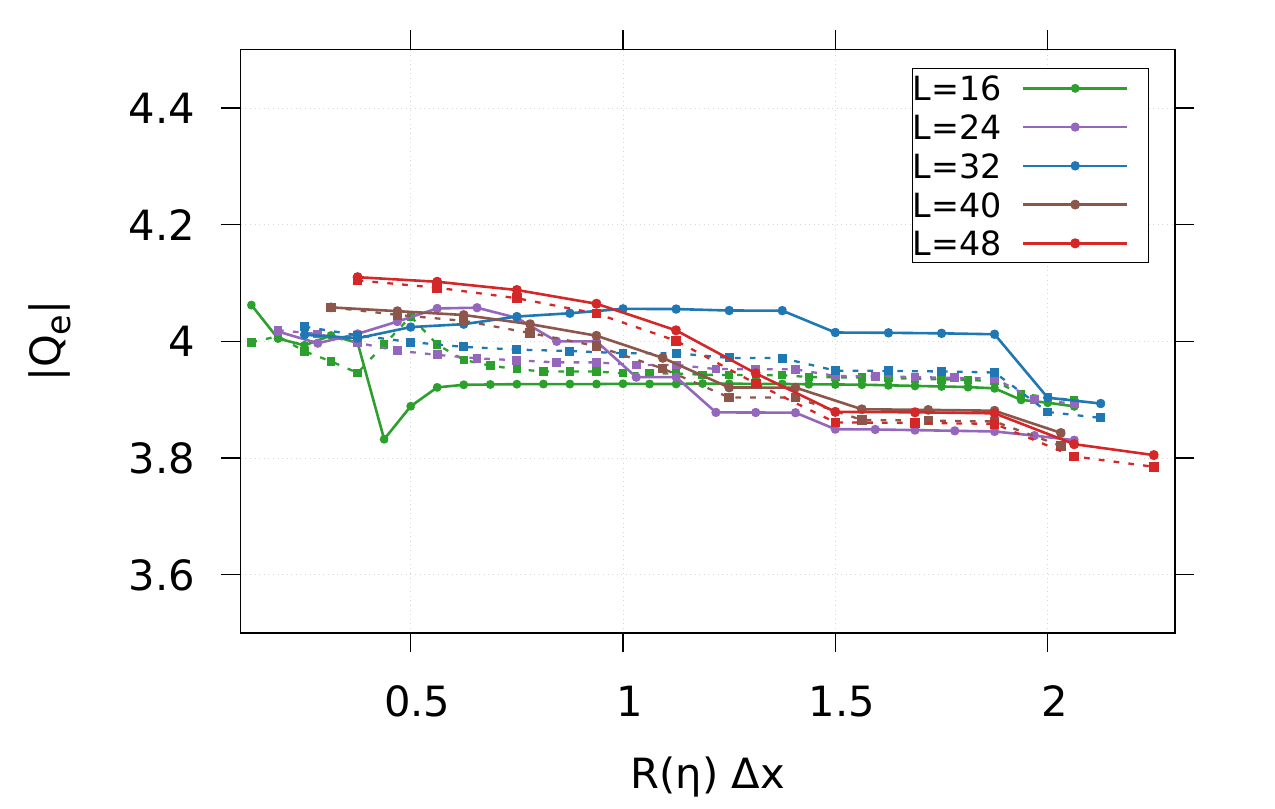}
\caption{Electric-charge retention for a dyon-antidyon pair initialized from static profiles with target charge magnitudes $|Q_e|=4$ and no added ambient charged excitations.
The horizontal axis is the physical lattice spacing $a_{\mathrm{phys}}=R\Delta x$.
The vertical axis shows $|Q_e(r)|$ at the outermost measurement surface for each object and output time.
Colors distinguish the comoving box sizes $L$ at fixed $N=256$.
Solid curves with circles show the dyon, and dashed curves with squares show the antidyon.
The core reference length is $r_{\mathrm{core}}=1/\sqrt{2}$.}
 \label{fig:static-charge-retention}
\end{figure}

Figure~\ref{fig:static-charge-retention} shows the electric-charge magnitude at the outermost measurement surface for each object and output time, using profiles measured out to $8r_{\mathrm{core}}$.
The measured magnitudes remain approximately $3.8$--$4.1$ over the displayed evolution, including physical lattice spacings larger than those reached in the formation simulations.
This test shows that the enclosed electric charge remains approximately constant in our $(3+1)$-dimensional lattice simulations, even when the physical lattice spacing exceeds the reference core size, $a_{\mathrm{phys}}>r_{\mathrm{core}}$.

\bibliography{reference}

\end{document}